\documentclass[onecolumn,preprint,preprintnumbers,amsmath,amssymb]{revtex4}

\usepackage{graphicx}
\usepackage{dcolumn}
\usepackage{bm}

\begin{document}
\title{Anomalous interaction between vortices and nanomagnets}

\author{Lars Egil Helseth}
\affiliation{Department of Chemistry and Biochemistry, Florida State University,
Tallahassee, Florida 32306-4390, USA}%

\begin{abstract}
We study a thin film system consisting of a superconducting and a magnetic film, where the superconductor 
contains a vortex and the magnetic film a nanomagnet. We find that if the magnetic film has planar anisotropy, 
the vortex induces a magnetization distribution, and its interaction with the nanomagnet crosses over from 
attractive to repulsive at short distances.

\end{abstract}

\pacs{Valid PACS appear here}
\maketitle

\section{Introduction}
The interaction between superconductivity and magnetism has been studied
for several decades. Systems composed of magnetic and
superconducting materials are of interest not only because they are model 
systems for the interplay of competing superconducting and magnetic order 
parameters, but also because of numerous possible 
applications\cite{Bulaevskii,Santiago,Jan,Jaccard,Daumens,Albrecht,Gardner,Goa,Helseth1,Helseth2}. 
Recently, the development of magnetic thin film technology has triggered a lot of 
interest in this field. Of particular importance has been the possibility of 
examining the interaction between vortices and 
nanomagnets, and it has been suggested that for isolated structures (e.g. a vortex close to a magnetic dot) such 
interactions are always attractive if the polarities of the vortex and the nanomagnet are the same (see e.g. 
Ref. \cite{Helseth1} and references therein).
However, when the nanomagnet is inserted in a soft magnetic environment, new phenomena may take place due to the fact that 
the vortex induces a magnetization distribution. It is our aim here to demonstrate that the interaction between the 
vortex and a 'hard' nanomagnet in a soft magnetic film with planar anisotropy crosses over from attractive to repulsive near the core 
of the vortex. 

To this end, consider a thin superconducting film of infinite extent, located at 
z=0 with thickness $d$ much smaller than the penetration depth of the
superconductor. The surface is covered by a uniaxial soft magnetic film with thickness comparable to or smaller than 
that of the superconducting film. That is, we assume that the magnetic film consists 
of surface charges separated from the superconductor by a very thin oxide layer
(of thickness t) to avoid spin diffusion and proximity effects. In 
general, the current density is a sum of the supercurrents and magnetically 
induced currents, which can be expressed through the generalized London 
equation as\cite{Helseth1,Helseth2} 
\begin{equation}
\mbox{\boldmath $\nabla \times J$} = -\frac{1}{\lambda ^{2}} \mbox{\boldmath $H$} + 
\frac{1}{\lambda ^{2} } V (\mbox{\boldmath $\rho$},\phi) \delta (z) \mbox{\boldmath $\hat{e}$}_{z} + 
\mbox{\boldmath $\nabla \times \nabla \times M$}_{V}\,\,\, ,
\label{LA}
\end{equation}
where $\mbox{\boldmath $J$}$ is the current density, $\lambda$ is the penetration depth, $\mbox{\boldmath $H$}$ is the 
magnetic field and $\mbox{\boldmath $\nabla$} \times \mbox{\boldmath $M$}_{V}$ 
is the magnetically induced current. Note that the magnetically induced currents are included as the last term on the 
right-hand side of Eq. \ref{LA}, and are therefore generated in the same plane as the vortex. This is justified since 
we assume that the thicknesses of both the superconducting and magnetic films are smaller than the penetration depth, and 
the magnetic film is located very close to the superconductor. We will here consider a Pearl vortex aligned in the z 
direction, with a source function $V(\rho) =(\Phi _{0} /\mu _{0}  )\delta (\rho)$\cite{Pearl}.
The magnetization $\mbox{\boldmath $M$}_{V}$ is the sum of two different contributions, $\mbox{\boldmath $M$}_{I}$ and $\mbox{\boldmath $M$}_{B}$. 

If the vortex has planar anisotropy (i.e. the magnetization vector is in the plane of the film in absence of magnetic fields), 
then the vortex induces a magnetization $\mbox{\boldmath $M$}_{I}$ in the magnetic film, where \cite{Helseth2} 
\begin{equation}
\mbox{\boldmath $M$}_{I} \approx M_{s}\left(1,\frac{H_{vz}}{H_{a}} \right) \,\,\, .
\label{tilt1}
\end{equation}
Here we denote $H_{v\rho}=\sqrt{H_{vx}^{2} +H_{vy}^{2}}$ and $H_{vz}$ as the in-plane and perpendicular components of the 
vortex field, and $M_{s}$ the saturation magnetization of the magnetic film. $H_{a}=M_{s} -2K_{u}/\mu_{0}M_{s}$ is the socalled anisotropy field, where 
$K_{u}$ is the anisotropy constant of the magnetic film. Here the easy axis is normal to the film surface, and we neglect 
the cubic anisotropy of the system. Upon using Eq. \ref{tilt1}, we neglect the contribution from the exchange energy, which 
can be justified for the magnetic field gradients considered here. 

In addition to the vortex-induced magnetization vector, there may also be 'hard' micromagnetic elements, e.g. prepatterned magnetization distributions or 
Bloch walls, that do not experience a 
significant change in their magnetization distribution $\mbox{\boldmath $M$}_{B}$ upon interaction 
with the vortex.

We are here interested in estimating the interaction between a circular symmetric micromagnetic element with perpendicular 
magnetization $M_{B}(\rho)\hat{\mbox{\boldmath $e$}}_{z}\delta (z)$ and the vortex. To this end, it should be noted 
that the magnetic induction in the magnetic film is given by 
$\mbox{\boldmath $B$}_{I} = \mu _{0}(\mbox{\boldmath $H$}_{v} +\mbox{\boldmath $M$}_{I})$. 
The interaction energy is given by
\begin{equation}
E = -\int_{-\infty}^{\infty} \int_{-\infty}^{\infty} \mbox{\boldmath $M$}_{B} \mbox{\boldmath $\cdot$}\mbox{\boldmath $B_{I}$} d^{2}\rho = -\mu _{0} \int_{-\infty}^{\infty} \int_{-\infty}^{\infty} M_{B}
\hat{\mbox{\boldmath $e$}}_{z} \mbox{\boldmath $\cdot$} \left( \mbox{\boldmath $H_{v}$} + \mbox{\boldmath $M_{I}$} \right) d ^{2}\rho \,\,\, ,
\label{energy}
\end{equation}
where the vortex is assumed to be displaced a distance $\rho _{0}$ from the origin of the nanomagnet.
It is clear that the radial component of the vortex-induced magnetization does not interact with $\mbox{\boldmath $M$}_{B}$, since the two vectors 
are perpendicular to each other. However, $\mbox{\boldmath $M$}_{B}$ does interact with the $z$-component of the vortex induced field, and we find
\begin{equation}
E =  -\mu _{0}\left( 1 + \frac{M_{s}}{H_{a}} \right) \int_{-\infty}^{\infty} \int_{-\infty}^{\infty} 
M_{B} H_{vz} d ^{2}\rho \,\,\, .
\label{interact}
\end{equation}
Using the method of Refs. \cite{Helseth1,Helseth2}, one finds in the cylindrically symmetric situation
\begin{equation}
E =  -\frac{\mu _{0}}{2\pi} \left( 1 + \frac{M_{s}}{H_{a}} \right)  \int_{0}^{\infty} kV(k)\frac{M_{B}(k) 
J_{0} (k\rho _{0}) }{1+2\lambda _{e}k -\alpha k^{2}} dk  \,\,\, , \,\,\, \alpha = \frac{d\lambda _{e}M_{s} }{H_{a} }  \,\,\, .     
\label{energyvm}
\end{equation}
The associated lateral force can be found by taking
the derivative with respect to $\rho _{0}$, resulting in
\begin{equation}
F = -\frac{\mu _{0}}{2\pi} \left( 1 + \frac{M_{s}}{H_{a}} \right) \int_{0}^{\infty} k^{2}V(k)\frac{M_{B}(k) 
J_{1} (k\rho _{0}) }{1+2\lambda _{e}k -\alpha k^{2}} dk     \,\,\, .
\label{forcevm}
\end{equation}
At small distances (large $k$) the force changes sign due to the magnetically induced currents. 
Let us now assume that the micromagnetic element is a magnetic dot so that $M_{B}(k)=M_0$ (i.e. that the magnetization 
distribution can be represented by a delta function), in order to gain some insight.
Then, for a Pearl vortex, the force becomes
\begin{equation}
F = -\frac{\Phi _{0}M_0}{2\pi} \left( 1 + \frac{M_{s}}{H_{a}} \right) \int_{0}^{\infty} k^{2}\frac{ 
J_{1} (k\rho _{0}) }{1+2\lambda _{e}k -\alpha k^{2}} dk     \,\,\, .
\end{equation}
When the distance is small (still larger than the coherence length $\xi$), we may approximate the force by
\begin{equation}
F \approx \frac{\Phi _{0} M_0}{2\pi \alpha \rho _{0}} \left( 1 + \frac{M_{s}}{H_{a}} \right) \,\,\, , \,\,\, \xi < \rho \ll \frac{M_{s}}{H_{a}} d \,\,\, .
\end{equation}
Surprisingly, we see that the force is repulsive, which is explained by the vortex-induced magnetization distribution. 
Thus, we have a crossover from attraction to repulsion close to the vortex core. 
For comparison, we note that in the approximation $\alpha \rightarrow 0$, the force is attractive, and 
at small distances given by
\begin{equation}
F \approx -\frac{\Phi _{0}M_0}{4\pi \lambda _{e}\rho _{0}^{2}} \left( 1 + \frac{M_{s}}{H_{a}} \right) \,\,\, , \,\,\, \xi < \rho \ll \lambda _{e} \,\,\, ,
\end{equation}
In both cases it is interesting to note that the force is enhanced by a factor $1 + M_{s}/H_{a}$ due to the soft 
magnetic film. Therefore it should be possible to tune the interaction by changing the uniaxial anisotropy of 
the magnetic film. For a magnetic dot without the magnetic film ($M_s=0$, $\alpha=0$) the force reduces to that 
previously studied \cite{Helseth1}. 
Finally it should be mentioned that attractive behavior also takes place at larger distances, $\rho _0 \gg \lambda _e$, where the term 
containing $\alpha$ becomes unimportant. 

In conclusion, we have shown that the interaction between a vortex and a hard nanomagnet in a soft magnetic film  
has a change in sign when the two elements get close to each other. This can be used to trap and expell the vortex by e.g. tuning the magnetization 
$M_s$ of the magnetic film. 

The author is grateful to T.M. Fischer for generous support.

\end{document}